\documentclass[aps,nofootinbib,superscriptaddress]{revtex4-2}
\usepackage{amsmath}   
\usepackage{amssymb}  
\usepackage{footnote}
\usepackage{geometry} 
\usepackage{amsmath}    
\usepackage{indentfirst}  
\usepackage{xcolor} 
\usepackage[title]{appendix}
\usepackage{graphicx} 
\usepackage{caption}  
\usepackage{subcaption}
\usepackage{booktabs}
\usepackage{siunitx}
\usepackage{physics}
\usepackage{hhline} 
\usepackage{booktabs}
\usepackage{float}
\usepackage{hyperref}

\begin{document}

\title{Modified Mukhanov-Sasaki equation and primordial perturbations in $\kappa$-deformed non-commutative space-time}

\author{Lei Yang}
\email{leiyang@hunnu.edu.cn}
\affiliation{Department of Physics, Key Laboratory of Low Dimensional Quantum Structures and Quantum Control of Ministry of Education, and Hunan Research Center of the Basic Discipline for Quantum Effects and Quantum Technologies, Hunan Normal University, Changsha, Hunan 410081, China} 

\author{Vishnu Rajagopal}
\email{vishnu@hunnu.edu.cn}
\affiliation{Department of Physics, Key Laboratory of Low Dimensional Quantum Structures and Quantum Control of Ministry of Education, and Hunan Research Center of the Basic Discipline for Quantum Effects and Quantum Technologies, Hunan Normal University, Changsha, Hunan 410081, China} 

\author{Puxun Wu}
\email{pxwu@hunnu.edu.cn}
\affiliation{Department of Physics, Key Laboratory of Low Dimensional Quantum Structures and Quantum Control of Ministry of Education, and Hunan Research Center of the Basic Discipline for Quantum Effects and Quantum Technologies, Hunan Normal University, Changsha, Hunan 410081, China} 
	
\begin{abstract}
We study the inflationary primordial perturbations in $\kappa$-Minkowski non-commutative space-time, a Lie-algebraic type deformation of canonical space-time motivated by quantum gravity scenarios. Employing the $\kappa$-deformed star product formalism, we construct the bilinear action for curvature perturbations and derive the $\kappa$-deformed Mukhanov–Sasaki equation and obtain the perturbative solutions. Further, we compute the primordial power spectrum and spectral index, showing that the leading order corrections to the power spectrum induces scale-dependent term proportional to $(\ln k)^2$. The spectral index also exhibits an explicit $\ln k$ dependence, which persists even when the slow-roll parameters are constant. We also perform a Bayesian MCMC analysis using ACT DR6 data and constrain the $\kappa$-deformation length scale to $\lambda=1.93^{+1.88}_{-1.33}\times 10^{-29}m$ at \(1\sigma\) CL, approximately four orders of magnitude larger than the Planck scale, demonstrating that the $\kappa$-deformed space-time offers a potential window into quantum gravity phenomenology through precision cosmology.
	\end{abstract}

    \maketitle
	
	\section{Introduction}

Einstein's general theory of relativity has been remarkably successful in describing the fundamental aspects of gravity at the macroscopic scale, and has been rigorously tested through numerous experiments and observations. However, it is inadequate for understanding the quantum nature of space-time structure. To comprehensively explore physics at the Planck scale, a complete quantum theory of gravity is essential, and various efforts at this direction are in progress. Although there is no complete theory of quantum gravity (QG), the Heisenberg's uncertainty principle implies that the inherent uncertainties at quantum scales suggest a discrete space-time structure, which can be effectively described by non-commutative (NC) space-time algebras~\cite{Snyder:1946qz, Doplicher:1994zv,Doplicher:1994tu}. NC geometry naturally incorporates a fundamental length scale relevant at Planck scale~\cite{noncommutativegeometry,Lizzi:2021bfq}, beyond which the space-time loses its operational meaning and consequently the NC space-time structures are replaced by the standard space-time manifolds. Recently, it has been shown that NC geometry can emerge naturally within perturbative QG at the Planck scale~\cite{Frob:2022ciq, Frob:2023vay}.
    
Various types of NC space-times have been rigorously constructed based on different physical motivations, and their implications have been extensively studied. Interestingly, certain NC space-times have been shown to emerge from  specific QG models. For example, the Moyal space-time, a well-developed canonical NC space-time \cite{Douglas:2001ba}, arises from the compactification of open strings in the presence of a constant background magnetic field, with its constant NC parameter being directly related to this magnetic field \cite{Seiberg:1999vs}. The $\kappa$-Minkowski space-time, which is a Lie-algebraic type NC space-time \cite{Arzano:2021hpg} whose symmetry algebra is defined by deformed Poincare algebras such as the $\kappa$-Poincare algebra \cite{Lukierski:1991pn,Lukierski:1992dt,Majid_1994},  has been shown to emerge naturally when loop quantisation techniques are applied to the algebra of gauge constraints of 3d Euclidean gravity with a positive cosmological constant \cite{Cianfrani:2016ogm}. Through the $\kappa$-Poincare algebra, the $\kappa$-Minkowski space-time can also  emerge as the underlying NC space-time of doubly special relativity (DSR), which introduces two observer-independent scales, characterising the kinematic structure of the underlying QG theory \cite{Amelino-Camelia:2002cqb,Kowalski-Glikman:2004fsz}. Recently, the $\kappa$-Minkowski space-time algebra has been found to appear upon introducing discontinuous paths in the path integral formulation \cite{Arzano:2024apl}. These characteristics of the $\kappa$-Minkowski space-time suggest that  it offers a compelling framework for describing the quantum nature of space-time structures.

Significant progress has been made in understanding the role of non-commutativity in cosmology, particularly regarding early-universe dynamics and observational signatures.~\footnote{The NC cosmological models have also been used to describe the accelerated expansion of the universe \cite{Kan:2019mxt,Oliveira-Neto:2021zqy,Abreu:2025wjv,Oliveira-Neto:2025bkj}. Interestingly in \cite{Das:2025pac}, the NC parameter is shown to act as an alternative source of accelerating expansion of the universe and this NC parameter has been constrained using the latest DESI DR1 datasets.} The primordial inflation in NC space-time has been studied extensively in the Moyal space-time, where non-commutativity is introduced either by replacing the point-wise multiplications of the inflationary action with the star product \cite{Lizzi:2002ib,Brandenberger:2002nq,Kim:2004kb,Xue:2007bb,Fang:2007ba,Calmet:2015fma,Kothari:2015xva,Tan:2018luz} or by employing NC quantum fields satisfying twisted statistics, compatible with the deformed Poincaré symmetry \cite{Akofor:2007fv,Koivisto:2010fk,Nautiyal:2013bwa,Jain:2014cpa}. These NC corrections are shown to introduce anisotropic signatures in the primordial spectrum. Recently, stringent bounds on the NC parameter of Moyal space-time have been obtained by performing a Bayesian analysis of inflationary models in Moyal space-time using the latest PLANCK PR4 datasets \cite{Gandhi:2026ktq}. Recently, inflationary dynamics within the NC framework has been rigorously investigated using the Hamiltonian formalism, where NC corrections are incorporated by deforming the Poisson bracket through the Moyal star product \cite{Sadeghi:2021egp,Rasouli:2022hnp,Nekouee:2022mvd,Kan:2022meh,NooriGashti:2025enx,Socorro:2024mwn}. Furthermore, inflationary dynamics and the late-time acceleration of the universe have been investigated within NC extensions of branch-cut quantum gravity \cite{Bodmann:2023hhx,Vasconcellos:2024ooo,ZenVasconcellos:2024wip} and LQG \cite{Diaz-Barron:2021yha,Diaz-Barron:2025bmi,Mohammadi:2024lgo} frameworks.
    
For the $\kappa$-deformed NC space-time, it has been utilized to remove the initial big bang singularity, allowing for a bounce behavior~\cite{Rajagopal:2025qyp}. Furthermore, scalar metric fluctuations in a $\kappa$-deformed Robertson-Walker space-time has been studied by constructing a non-local scalar field theory in the deformed space-time by generalizing the $\kappa$-Minkowski field theory to Robertson-Walker space-time~\cite{Kim:2005tf}. It has been found that the $\kappa$-deformation induces non-trivial time-dependent and momentum-dependent corrections to the power spectrum. Recently, we investigated primordial perturbations in $\kappa$-deformed space-time by assuming that the perturbation modes follow the standard Mukhanov-Sasaki equation. We introduced non-commutative (NC) effects in the two-point correlator of the primordial power spectrum through the $\kappa$-deformed oscillator algebra, leading to the discovery of non-trivial scale-dependent modifications in both the scalar and tensor power spectra~\cite{Rajagopal:2025vbs}. However, it has been shown that perturbation modes in Moyal non-commutative space-time satisfy a modified Mukhanov-Sasaki equation \cite{Lizzi:2002ib,Brandenberger:2002nq,Kim:2004kb,Xue:2007bb,Fang:2007ba,Calmet:2015fma,Kothari:2015xva,Tan:2018luz}. Additionally, the non-commutative fields in $\kappa$-deformed space-time exhibit modified dispersion relations that are compatible with the $\kappa$-Poincare algebra \cite{Lukierski:1991pn,Lukierski:1992dt}, which should result in deformation of the corresponding equations of motion. Therefore, to consistently study the $\kappa$-deformation of primordial perturbations, it is essential to derive a modified Mukhanov-Sasaki equation from first principles that aligns with the symmetry algebra of $\kappa$-Minkowski space-time.

In this work, we study primordial fluctuations in $\kappa$-deformed space-time by constructing a $\kappa$-deformed bilinear action for the curvature perturbation using the star product formalism, which is compatible with the deformed Poincare algebra of $\kappa$-Minkowski space-time, as discussed in \cite{Meljanac:2006ui,Kresic-Juric:2007vgu,Meljanac:2007xb}. The introduction of the star product results in higher-derivative terms within the action, which  modifies the  Mukhanov-Sasaki equation. To avoid the possible ghost instabilities associated with  these higher-order derivatives, we adopt an effective field theory framework, allowing us to solve perturbatively the deformed Mukhanov-Sasaki equation. Our analysis reveals that the leading-order correction to the scalar power spectrum  contains $(\ln k)^2$ terms, leading to a scale-dependent enhancement of the scalar spectral index. In addition, we also perform a comprehensive Bayesian analysis using ACT DR6 data \cite{AtacamaCosmologyTelescope:2025blo} and constrain the NC parameters of $\kappa$-deformed space-time.  

This paper is organised in the following way. In Section.(\ref{sec2}), we construct the $\kappa$-deformed Mukhanov-Sasaki equation using the star product formalism. In Section.(\ref{sec3}), we obtain the perturbative solutions and calculate the power spectrum for curvature perturbation, under $\kappa$-deformation. In Section.(\ref{sec4}), we constrain the NC parameters by performing MCMC analysis using ACT DR6 data. Finally, in Section.(\ref{sec5}), we summarise our results and provide the concluding remarks.

\section{$\kappa$-modified Mukhanov-Sasaki equation}\label{sec2}
	
In general, field theories in NC space-times are studied by promoting the point-wise multiplication to a star multiplication that is compatible with the deformed  Poincaré algebra of the underlying NC space-time. We extend this procedure to $\kappa$-deformed space-time by employing the $\kappa$-deformed star product formalism discussed in \cite{Meljanac:2006ui,Kresic-Juric:2007vgu,Meljanac:2007xb}. Utilizing this $\kappa$-deformed star product formalism, we construct the bilinear action for curvature perturbations in $\kappa$-deformed space-time, which leads to the derivation of the modified Mukhanov-Sasaki equation.

The $\kappa$-Minkowski space-time algebra is defined as \cite{Kresic-Juric:2007vgu}
\begin{equation}\label{kappa}
		\left[\hat{x}_{\mu}, \hat{x}_{\nu}\right]=i\left(a_{\mu}\hat{x}_{\nu} - a_{\nu}\hat{x}_{\mu}\right)
	\end{equation}
Here, $a_{\mu}$ is defined as $a_{\mu}=(\lambda,0,0,0)$, where $\lambda$ is the $\kappa$-deformation parameter and has the length dimension. From this, it can be seen that the $\kappa$-Minkowski space-time preserves the rotational symmetry and deforms the Lorentz sector, leading to a deformed Poincare algebra. In general the symmetry of $\kappa$-Minkowski space-time is defined in terms of the $\kappa$-Poincare algebra, where both the algebra and co-algebra sectors are deformed \cite{Lukierski:1991pn,Lukierski:1992dt,Majid_1994}. However, the symmetry algebra of $\kappa$-Minkowski can also be realised alternatively using an undeformed $\kappa$-Poincare algebra \cite{Meljanac:2006ui,Kresic-Juric:2007vgu,Meljanac:2007xb}, where the standard form of the Poincare algebra is retained, by re-defining the explicit form of Lorentz and translation generators, respectively. This has been achieved by realising the $\kappa$-deformed NC space-time coordinates as function of the commutative space-time coordinate $x_{\mu}$ and its conjugate $p_{\mu}$, using a realisation function $\varphi^{\alpha} \! _{\mu} (p)$, in the following manner
\begin{equation}
\hat{x}_{\mu}=x_{\alpha}\varphi^{\alpha} \! _{\mu} (p).
\end{equation}
Demanding the consistency of this realisation with the $\kappa$-Minkowski space-time algebra given in Eq.(\ref{kappa}), we obtain a differential equation for $\varphi^{\alpha} \! _{\mu} (p)$ as, $\frac{\partial \varphi^{\alpha} \! _{\mu}}{\partial p^{\beta}}\varphi^{\beta} \! _{\nu}-\frac{\partial 
\varphi^{\alpha} \! _{\nu}}{\partial p^{\beta}}\varphi^{\beta} \! _{\mu} =
a_{\mu}\varphi^{\alpha} \! _{\nu}-a_{\nu}\varphi^{\alpha} \! _{\mu}$. Solving this equation perturbatively, up to first order in the $\kappa$-deformation parameter $a$, we get \cite{Harikumar:2012zi}
\begin{equation}\label{phi}
\varphi^{\alpha} \! _{\mu}=\delta^{\alpha} _{\mu}\left(1+\alpha(a\cdot p)\right)+\beta a^{\alpha} p_{\mu}+\gamma 
p^{\alpha} a_{\mu},
\end{equation}
where $\alpha,\beta$ and $\gamma$ are dimensionless parameters that appear in the realisation, \textcolor{red}{such that they should satisfy the constraints, $\gamma-\alpha=1,~\beta\in\mathbb{R}$}. Using the above obtained realisation, a NC function in $\kappa$-deformed space-time is realised in terms of commutative counterparts as \cite{Harikumar:2012zi}
\begin{equation}
\hat{f}=f(x)+\alpha\left(x\cdot \frac{\partial f}{\partial x}\right)\left(a\cdot p\right)+\beta\left(a\cdot x\right)\left(\frac{\partial f}{\partial x}\cdot p\right)+\gamma\left(a\cdot \frac{\partial f}{\partial x}\right)\left(x\cdot p\right)
\end{equation}
In general, the NC effects in field theories are incorporated by promoting the point-wise multiplications in the action with a star multiplication, i.e., $f(x)\cdot g(x)\to f(x)\star g(x)$. Since there exists an isomorphism between the NC algebra (generated by NC coordinates) and the star product algebra (generated by commutative coordinates), we can write them as, $f(x)\star g(x)= \hat{f}(\hat{x})\cdot\hat{g}(\hat{x})\triangleright 1$. Here, $\triangleright$ is a mapping between the Heisenberg algebra and the commutative space-time algebra, whose action on $f(x)$ is defined as $x_{\mu}\triangleright f(x)=x_{\mu}f(x)$ and $p_{\mu}\triangleright f(x)=i\partial_{\mu}f(x)$. Similarly, the action of $\triangleright$ on the unit element is defined as, $x_{\mu}\triangleright 1=x_{\mu}$ and $p_{\mu}\triangleright 1=0$. Now, substituting the explicit form of $\hat{f}$ and $\hat{g}$ in $f(x)\star g(x)= \hat{f}(\hat{x})\cdot\hat{g}(\hat{x})\triangleright 1$ and using the above defined action of $\triangleright$, we obtain the $\kappa$-deformed star product between the two functions ${f}(x)$ and ${g}(x)$, valid up to first order in the deformation parameter $\lambda$, as \cite{Gupta:2013ata}
\begin{equation}\label{star1}
f(x)\star g(x)=f(x)g(x)+i\alpha\left(x\cdot \frac{\partial f}{\partial x}\right)\left(a\cdot \frac{\partial g}{\partial x}\right)+i\beta\left(a\cdot x\right)\left(\frac{\partial f}{\partial x}\cdot \frac{\partial g}{\partial x}\right)+i\gamma\left(a\cdot \frac{\partial f}{\partial x}\right)\left(x\cdot \frac{\partial g}{\partial x}\right)
\end{equation}
This star product encodes the effects of non-commutativity and plays a crucial role in constructing the actions in NC space-times. From the first order correction terms of Eq.(\ref{star1}), it is evident that the NC induces higher order derivatives in the action and the subsequent equations of motions obtained from it. These correction terms are shown to exhibit some interesting features, such as non-locality and mixing of UV-IR divergences \cite{Grosse:2005iz}. As mentioned earlier, in this study we try to explore how these correction terms affect the propagation of curvature perturbations during  inflation. To begin this discussion, we consider the standard bilinear action for the curvature perturbation \cite{Baumann:2022mni}
	\begin{equation}\label{dot-action}
		S^{(2)}=\frac{1}{2} \int d^{4} x ~\sqrt{-g} ~\frac{\dot{\phi}^{2}}{H^{2}} ~g^{\mu \nu}\left(\partial_{\mu} \mathcal{R}\cdot\partial_{\nu} \mathcal{R}\right),
	\end{equation}
where $\phi(x)$ is the inflaton field, $\mathcal{R}$ is the curvature perturbation and $g_{\mu \nu}$ is the background FLRW metric. Now we generalise this action into $\kappa$-deformed space-time, by promoting the standard point-wise multiplication into the star product
	\begin{equation}\label{star-action}
		S^{(2)}_{\lambda}=\frac{1}{2} \int d^{4} x~\sqrt{-g}~\frac{\dot{\phi}^{2}}{H^{2}} ~g^{\mu \nu}\left(\partial_{\mu} \mathcal{R} \star \partial_{\nu} \mathcal{R}\right),	
    \end{equation}
We compute the star product between the covariant derivative of curvature perturbations in the above action, by replacing $f(x)$ and $g(x)$ of Eq.(\ref{star1}), with $\partial_{\mu} \mathcal{R}$ and $\partial_{\nu} \mathcal{R}$, respectively.
	\begin{equation}\label{star-r}
		\partial_{\mu} \mathcal{R} \star \partial_{\nu} \mathcal{R}=\partial_{\mu} \mathcal{R} \partial_{\nu} \mathcal{R}+i a \alpha x^{\rho} \partial_{\rho} \partial_{\mu} \mathcal{R} \partial_{0} \partial_{\nu} \mathcal{R}+i a \beta x^{0} \partial_{\rho} \partial_{\mu} \mathcal{R} \partial^{\rho} \partial_{\nu} \mathcal{R}+i a \gamma x^{\rho} \partial_{0} \partial_{\mu} \mathcal{R} \partial_{\rho} \partial_{\nu} \mathcal{R}.
	\end{equation}
As mentioned earlier, here the star product induces higher derivative curvature perturbation terms in the action and in the resulting equations of motion. Substituting Eq.(\ref{star-r}) in Eq.(\ref{star-action}), we evaluate the star product and re-defining the curvature perturbation via $\varphi=z{\cal R}$, we obtain the $\kappa$-deformed bilinear action as
  \begin{equation}
		S^{(2)}_{\lambda}=\frac{1}{2} \int d\eta ~d^{3} x\left(\left(\varphi^{\prime}\right)^{2}-\left(\partial_{i} \varphi\right)^{2}+\frac{z^{\prime \prime}}{z} \varphi^{2}+\frac{i\lambda}{a} \eta\left(c_{1}\left(\varphi^{\prime \prime}\right)^{2}-c_{2}\left(\partial_{i} \varphi^{\prime}\right)^{2}+c_{3}\left(\partial_{i} \partial_{j} \varphi\right)^{2}\right)\right).
	\end{equation}
where $z = a \frac{\dot{\phi}}{H}$ and $\prime$ denotes the derivative with respect to conformal time $\eta$. By varying this action and transforming it into Fourier space, we obtain the modified Mukhanov-Sasaki equation at first order in the $\kappa$-deformation parameter.
	\begin{equation}\label{eq:S_2new}
		\varphi_{k}^{\prime \prime}+\left(k^{2}-\frac{z^{\prime \prime}}{z}\right) \varphi_{k}+i \lambda \left(c_{1} \partial_{\eta}^{2}\left(\frac{\eta}{a} \varphi_{k}^{\prime \prime}\right)+c_{2} k^{2} \partial_{\eta}\left(\frac{\eta}{a} \varphi_{k}^{\prime}\right)+c_3\frac{\eta}{a} k^{4} \varphi_{k}\right)=0 ,
	\end{equation}
Note that in the above expressions, we introduce the dimensionless coefficients $c_1=\alpha+\beta+\gamma,c_2=\alpha+2\beta+\gamma,c_3=\beta$, for simplification. We find that these three coefficients satisfy 
\begin{equation}
c_1-c_2+c_3=0, 
\end{equation} 
suggesting that only two of them are independent. 
  
\section{$\kappa$-deformed Primordial Power Spectrum}\label{sec3}

The first order NC corrections of the $\kappa$-deformed Mukhanov-Sasaki equation, obtained in Eq.(\ref{eq:S_2new}), through the $\kappa$-deformed star product, induces higher derivative terms such as $\partial_{\eta}^{2}\left(\eta \varphi_{k}^{\prime \prime}\right)$, which gives a fourth-order time derivative $\varphi_{k}^{\prime \prime\prime\prime}$. Such higher-order time derivatives could generally lead to the Ostrogradsky instability and as a result it would exhibit ghost states. In this regard, we adopt the techniques of effective field theory (EFT) \cite{Cheung_2008,burgess2021intro}, where the higher order terms can be eliminated at the level of equations of motion, valid upto first order in the $\kappa$-deformation parameter, by using the commutative equations of motion. In this way, all fourth-order and third-order terms of the $\kappa$-deformed Mukhanov-Saski equation are completely replaced by the lower order derivative terms. The resulting equations on motion would be devoid of higher time derivatives, and the Ostrogradsky instability as well as the emergence of ghost states can also be avoided. We begin this analysis by considering the standard Mukhanov-Sasaki equation, which can be obtained by setting $\lambda$ to be zero in Eq.(\ref{eq:S_2new}) 
	\begin{equation}
		\varphi_k^{\prime \prime}+\left(k^{2}-\frac{z^{\prime \prime}}{z}\right) \varphi_k=0.
	\end{equation} 
    Defining $\mathcal{K}(\eta)=k^{2}-\frac{z^{\prime \prime}}{z}$, the second derivative terms can be expressed as
	\begin{equation}\label{second}
		\varphi_k^{\prime \prime}=-\mathcal{K}(\eta) \varphi_k.
	\end{equation}
	Upon further differentiating Eq.(\ref{second}), we obtain the third- and fourth-order derivatives as
	\begin{equation}\label{third-fourth}
    \begin{split}
		\varphi_k^{\prime \prime \prime} &= -\mathcal{K}^{\prime}(\eta) \varphi_k-\mathcal{K}(\eta) \varphi_k^{\prime},\\
		\varphi_k^{\prime \prime \prime \prime} &=\left[\mathcal{K}^{2}(\eta)-\mathcal{K}^{\prime \prime}(\eta)\right] \varphi_k-2 \mathcal{K}^{\prime}(\eta) \varphi_k^{\prime}.
    \end{split}    
	\end{equation}
During slow roll inflation, $H$ varies slowly, and hence we take $\frac{z''}{z}\simeq\frac{2}{\eta^2}$, as $a\simeq -1/H\eta$. Now substituting Eq.(\ref{third-fourth}) into Eq.(\ref{eq:S_2new}), we obtain the $\kappa$-deformed corrections of Mukhanov-Sasaki equation in lower-order derivative terms
	\begin{eqnarray}\label{eft-ms}
		\varphi_k^{\prime \prime}+\mathcal{K} \varphi_k - i \lambda H \Big(c_{1}\left(\left(\eta^2\left(\mathcal{K}^{2}- \mathcal{K}^{\prime \prime}\right) -4\eta\mathcal{K}^{\prime}-2\mathcal{K}\right)\varphi_k -2\eta\left(2\mathcal{K}+\eta\mathcal{K}^{\prime}\right)\varphi_k'\right)     \nonumber \\ - c_2k^2\eta\left(\mathcal{K}\varphi_k-2\varphi_k'\right)+c_{3} k^{4}\eta^2\varphi_k \Big)=0.
	\end{eqnarray}
Now the $\kappa$-deformed Mukhanov-Sasaki equation, valid up to first in $\lambda$, contains only the lower derivative terms, as discussed earlier. Hence, the resulting deformed solution of this equation would not exhibit ghost states. In order to analyse the effects of $\kappa$-deformation on primordial perturbations, we need to obtain the solutions to $\kappa$-deformed Mukhanov-Sasaki equation given in Eq.(\ref{eft-ms}). We proceed this by obtaining the solutions in the super-horizon and sub-horizon limits, separately, using the perturbation method, valid up to first order in the $\kappa$-deformation parameter. The unknown coefficients of the solutions in sub-horizon and super-horizon limits are determined by matching them at the horizon crossing. 
    
In the sub-horizon limit $k|\eta| \gg 1$, we have $\mathcal{K} \simeq k^{2}, \mathcal{K}^{\prime}\simeq 0,~\mathcal{K}^{\prime \prime} \simeq 0$, which lead  the $\kappa$-deformed Mukhanov-Sasaki equation in Eq.(\ref{eft-ms}) to become
	\begin{equation}
		\varphi_k^{\prime \prime}+k^{2} \varphi_k \simeq  i\lambda H \left(2(c_2-2c_1)\eta k^2 \varphi_k' - 2c_1k^2\varphi_k\right)
	\end{equation}
Note that the coefficients of $k^4$ dependent correction term of the above equation vanishes since $c_1-c_2+c_3=0$. In order to solve this, we consider the perturbation method where $\varphi_k=\varphi_{k}^{(0)}+\lambda \varphi_{k}^{(1)}$. Thus, the unperturbed equation becomes $\varphi_{k}^{(0)\prime \prime}+k^{2} \varphi_{k}^{(0)}=0$, whose solution is given by the standard the Bunch–Davies vacuum condition as $\varphi_{k}^{(0)}(\eta)=\frac{e^{-i k \eta}}{\sqrt{2 k}}$ \cite{Baumann:2022mni}. On the other hand, the perturbed equation takes the form
	\begin{equation}
		\varphi_k^{(1)''} + k^2 \varphi_{k}^{(1)} = i\lambda H \left(2(c_2-2c_1)\eta k^2 \varphi_{k}^{(0)'} - 2c_1k^2\varphi_{k}^{(0)}\right).
	\end{equation}
Substituting $\varphi_{k}^{(0)}(\eta)=\frac{e^{-i k \eta}}{\sqrt{2 k}}$ in the LHS and solving the resulting inhomogeneous differential equation, we obtain the perturbative solution in the sub-horizon limit, valid upto first order in $\lambda$, as
	\begin{equation}\label{ms-sub}
		\varphi_{k}^{\text{sub}}(\eta)\simeq\frac{e^{-i k \eta}}{\sqrt{2 k}}+\lambda \left[ A e^{ik \eta} + B e^{-ik \eta} + \frac{H}{\sqrt{2k}} \left( \frac{i(c_2-2c_1)k^2\eta^2}{2} + \frac{c_2k\eta}{2} - \frac{ic_2}{4} \right) e^{-ik \eta}\right] .
	\end{equation}
In the super-horizon limit $k|\eta| \ll 1$, the $\mathcal{K}(\eta)$ and their derivatives becomes $\mathcal{K} \simeq-\frac{2}{\eta^{2}},~\mathcal{K}^{\prime} \simeq \frac{4}{\eta^{3}}$, and $\mathcal{K}^{\prime \prime} \simeq-\frac{12}{\eta^{4}}$, respectively. Using this in Eq.(\ref{eft-ms}), we get the $\kappa$-deformed Mukhanov-Sasaki equation, in the super-horizon limit as
	\begin{equation}\label{ms-super}
		\varphi_k^{\prime \prime}-\frac{2}{\eta^{2}} \varphi_k \simeq \frac{i 4\lambda H c_{1}}{\eta^2} \varphi_k
	\end{equation}
	Employing the trial power solution method $\varphi_k(\eta)=\eta^{p}$, Eq.(\ref{ms-super}), can be re-casted as
	\begin{equation}\label{p}
		p^{2}-p-(2+4 i \lambda H c_{1}) = 0  
	\end{equation}
We use the perturbation method to solve this equation, considering $p$, valid up to first order in $\lambda$, as $p=p_{0}+\lambda p_{1}$. Using this in Eq.(\ref{p}), we get the unperturbed part as $p_{0}^{2}-p_{0}-2=0$ and choose the frozen mode solution, i.e., $p_0=-1$. Now by substituting $p=-1+\lambda p_1$ in Eq.(\ref{p}), and keeping the terms valid upto first order in $\lambda$, we find $p_{1}=-\frac{4 i c_{1} H}{3}$. From this get the explicit of curvature perturbation $\varphi_k$ in the super-horizon limit, valid up to first order in $\lambda$, as 
	\begin{equation}\label{eq:longsolution}
		\varphi_{k}^{\text {super}}(\eta) \simeq C \eta^{-1}\left(1 - \frac{ 4i \lambda Hc_{1}}{3} \ln |\eta|\right).  
	\end{equation}
At the horizon crossing $k\eta=-1$, the sub-horizon mode $\varphi_k^{\text{sub}}$, obtained in Eq.(\ref{ms-sub}), becomes equal to the super-horizon mode $\varphi_k^{\text{super}}$, obtained in Eq.(\ref{ms-super}), and from this matching, we determine the coefficients $A,B,C$. Substituting these coefficients, we obtain the explicit form of $|\varphi_k|^2$ at horizon crossing as
	\begin{equation}\label{phi2}  
	 |\varphi_k|^2=\frac{1}{2k}\left(1+\frac{16}{9} \lambda^{2} c_{1}^{2} H^{2} \ln ^{2} k\right), 
	\end{equation}
which indicates that the NC correction to the super-horizon evolution of $|\varphi_k|^2$ introduces a logarithmic dependence on the co-moving wave number. The $k^2$-dependent correction term of the sub-horizon mode gets absorbed into the coefficients by matching the modes at horizon crossing. Furthermore, we find that the NC correction in Eq.(\ref{phi2}) only depends on the coefficient $c_1$. Using Eq.(\ref{phi2}) in relations $\mathcal{R}=\frac{\varphi_k}{z}$ and $z^2=2a^2\varepsilon$, we get
    \begin{equation}\label{Pnew}
		|\mathcal{R}_k|^2=\frac{H^{2}}{4k^3 \varepsilon}\left(1+\frac{16}{9} \lambda^{2} c_{1}^{2} H^{2} \ln ^{2} k\right),
	\end{equation}
where $\varepsilon$ is the slow roll parameter, defined as $\varepsilon=-\frac{\dot{H}}{H^{2}}$ \cite{Baumann:2022mni}. Substituting Eq.(\ref{Pnew}) in the standard definition for the scalar power spectrum $\mathcal{P}_{\mathcal{R}}=\frac{k^3}{2\pi^2}|\mathcal{R}_k|^2$ \cite{Baumann:2022mni}, we get the $\kappa$-deformed power spectrum for the scalar perturbation as 
    \begin{equation}\label{Pnew1}
		\mathcal{P}_{\mathcal{R}}=\frac{H^{2}}{8\pi^2 \varepsilon}\left(1+\frac{16}{9} \lambda^{2} c_{1}^{2} H^{2} \ln ^{2} k\right).
	\end{equation}
The primordial scalar power spectrum therefore acquires a scale- and logarithmic-dependent correction induced by spacetime NC, and this corrections depends only $c_1$ associated with super-horizon evolution shown in Eq.(\ref{phi2}). This expression shows that the non-commutativity introduces a quadratic enhancement at leading order, which grows with the logarithm of the co-moving wave number, providing a distinct observational signature of quantum space-time structures.
    
Although the higher derivative operators in the leading order corrections of the star product formalism introduces $k^2$ and $k^4$ terms in the Mukhanov-Sasaki equation, the resulting power spectrum obtained at the horizon crossing regime does not appear as explicit power-law corrections, as discussed above. Similarly, the higher dimensional operators of the extended effective field theory of inflation (EEFToI) introduce $k^6$, $k^4$, and $k^2$ correction terms in the Mukhanov-Sasaki equation and generate a correction that appears as a multiplicative factor to standard power spectrum, rather than a power-law type \cite{Ashoorioon:2018uey}. Thus, in both of these frameworks, where EFT techniques have been employed, the higher-derivative corrections do not appear as explicit power-law corrections to the final power spectrum.   
    
The $\kappa$-deformed power spectrum obtained in Eq.(\ref{Pnew1}) satisfies both parity and rotational invariance. This isotropy is a direct consequence of the $\kappa$-Minkowski space-time structure (see Eq.(\ref{kappa})), where the deformation parameter preserves the rotational symmetry while deforming the Lorentz sector. Therefore, the power spectrum is direction independent in the $\kappa$-deformed space-time and as a result, the angular power spectrum retains its diagonal form, and the statistical isotropy of the CMB is maintained. In contrast, the NC parameter of Moyal space-time explicitly breaks the rotational symmetry, and consequently the power spectrum also breaks the rotational invariance, although parity is preserved \cite{Akofor:2007fv}. Hence, the Moyal deformed power spectrum becomes direction dependent, leading to off-diagonal correlations in the angular power spectrum, resulting in statistical anisotropy. Therefore, this difference in the symmetry properties between these two NC space-times lead to distinct observational signatures in the CMB spectrum.

The absence of statistical anisotropy in the $\kappa$-deformed space-time provides a viable alternative to anisotropic NC models, with observational signatures appearing primarily through the scale-dependent running of the spectral index. Hence in order to obtain the expression of $\kappa$-deformed spectral index, we consider the general definition of the power spectrum $\mathcal{P}_S=A_sk^{n_s-1}$, where $A_s$ is the amplitude, and from this the scalar spectral index $n_s$ is defined as $n_s-1=\frac{d \ln \mathcal{P}_S}{d \ln k}$ \cite{Baumann:2022mni}. Substituting Eq.(\ref{Pnew}) in this definition, we get the expression for $\kappa$-deformed spectral index as
	\begin{equation}\label{scalar-index}
		n_{s}=1-2 \varepsilon-\delta+\frac{32}{9} \lambda^{2} c_{1}^{2} H^{2} \ln k(1 - \varepsilon \ln k),
	\end{equation}
where $\delta=\frac{\dot{\varepsilon}}{H \varepsilon}$ is the second slow-roll parameter \cite{Baumann:2022mni}. Here the scalar power spectrum in $\kappa$-deformed space-time is inherently scale dependent, as the leading order correction term contains the $\ln k$ term, in comparison to \cite{Rajagopal:2025vbs}, where the scale dependence appears from the running of slow-roll parameters and Hubble parameter. Interestingly, here this scale dependence persists even when the slow-roll parameters remain constant, in contrast to the standard slow-roll inflation where the scale dependence of the spectral index arises solely from the running of the slow-roll parameters. This characteristic scale-dependent spectral index provides a distinctive observational signature of $\kappa$-deformed space-time, detectable through precision measurements of the CMB spectral index across different scales.

\section{Observational constraints}\label{sec4}

Employing the modified scalar spectral index derived in Eq. (\ref{scalar-index}), we perform a comprehensive Bayesian Markov Chain Monte Carlo (MCMC) analysis using the ACT DR6 data \cite{AtacamaCosmologyTelescope:2025blo}, and constrain the $\kappa$-deformation scale $\lambda$ and the dimensionless NC parameter $c_1$, enabling us to assess the viability of this NC inflationary framework against current observations and establishing quantitative bounds on the scale of quantum space-time structures.

The binned ACT DR6 CMB power spectrum data \cite{AtacamaCosmologyTelescope:2025blo} provide the multipole moments \(\ell\), the angular power spectrum \(D_\ell\), and the associated uncertainties \(\sigma_{D_\ell}\) Using the $N=84$ data points from this dataset, corresponding to adjacent multipole bins spanning $l=2$ to $l=4152$, we extract values of the observed spectral index and its uncertainties.

We perform MCMC analysis using the Python \texttt{emcee} package \cite{Foreman-Mackey:2012any}. The log-likelihood function for this sampling is constructed as
\[
\ln \mathcal{L}(\theta) = -\frac{1}{2} \sum_{i=1}^N \left( \frac{n_s^{\text{model}}(k_i, \theta) - n_s^{\text{obs}}(k_i)}{\sigma_i} \right)^2,
\]
where \(\theta = \{\epsilon, \delta, c_1,\lambda\}\) are the free parameters and $n_s^{\text{model}}(k_i,\theta)$ is the $\kappa$-deformed scalar spectral index derived in Eq.(\ref{scalar-index}). The observed spectral index $n_s^{\text{obs}}(k_i)$ and its uncertainties $\sigma_i$ are extracted from \cite{AtacamaCosmologyTelescope:2025blo}, as discussed above. 

The choice of prior distributions for the model parameters are given in Table.(\ref{tab:prior}). Here we have imposed Gaussian priors for the slow-roll parameters $\varepsilon,\delta$ and the NC parameter $c_1$, and uniform log prior for the $\kappa$-deformation scale $\lambda$. Note that in this analysis, we set the Hubble parameter as $H=10^{14}GeV$. 
\begin{table}[htb]
\centering
\begin{tabular}{c @{\hspace{0.6cm}} c}
\hline\\\
\textbf{Parameter} & \textbf{Prior range} \\[0.2cm]
\hline\\
$\log_{10}(\lambda)$ & $\mathcal{U}[-34, -26]$ \\[0.2cm]
$\varepsilon$ & $\mathcal{U}[0.001,0.0005]$ \\[0.2cm]
$\delta$ & $\mathcal{U}[0.007, 0.011]$ \\[0.2cm]
$c_1$ & $\mathcal{U}[0.0005, 0.0015]$ \\[0.2cm]
\hline
\end{tabular}
\caption{\small Prior values of the model parameters}
\label{tab:prior}
\end{table}

\begin{figure}
    \centering
    \includegraphics[width=0.9\linewidth]{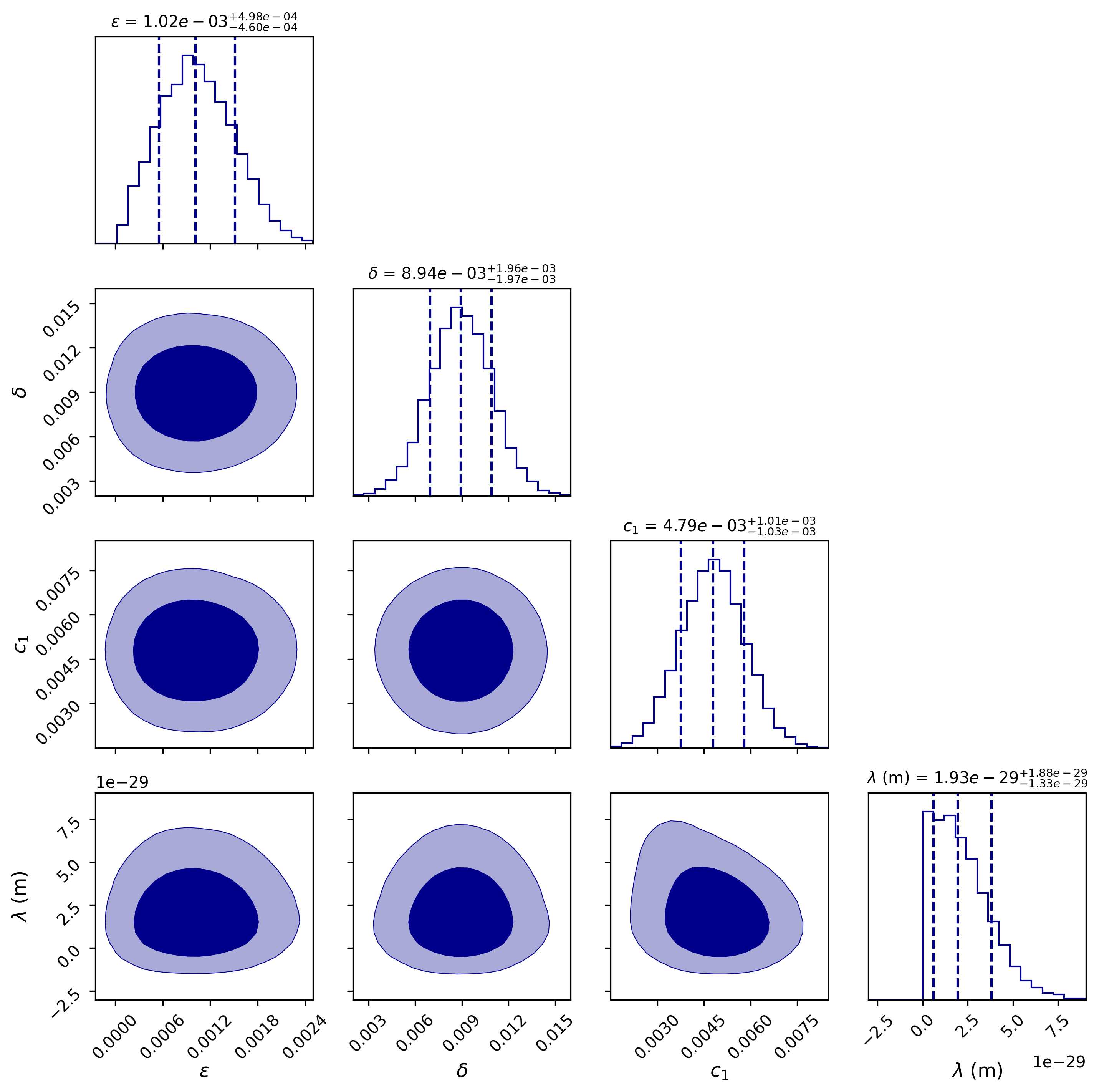}
    \caption{Marginalised contour and posterior distribution of the model parameters $\varepsilon,\delta,c_1\lambda$, as obtained from the ACT DR6 dataset}
    \label{fig:mcmc}
\end{figure}

\begin{table*}[htbp]
\centering
\begin{tabular}{c @{\hspace{0.6cm}} c}
\hline\\
\textbf{Parameter} & \textbf{Marginalised values at $1\sigma$ CL} \\[0.2cm]
\hline\\
$\lambda~(m)$ & $1.93^{+1.88}_{-1.33}\times 10^{-29}$ \\[0.2cm]
$\varepsilon$ & $1.02^{+0.498}_{-0.460}\times 10^{-3}$ \\[0.2cm]
$\delta$ & $8.94^{+1.96}_{-1.97}\times 10^{-3}$ \\[0.2cm]
$c_1$ & $4.79^{+1.01}_{-1.03}\times 10^{-3}$ \\[0.2cm]
\hline
\end{tabular}
\caption{\small Marginalised values of the slow-roll parameters, NC length scale and the dimensionless NC parameter at $1\sigma$ CL}
\label{tab:posterior}
\end{table*}

The marginalised posterior distributions of the MCMC analysis, shown in Fig.(\ref{fig:mcmc}), reveal well-constrained parameters for the $\kappa$-deformed inflationary model constructed in this study. The posterior of the slow-roll parameters exhibit nearly Gaussian distributions and the values obtained at $1\sigma$ confidence level, $\varepsilon=1.02^{+0.498}_{-0.460}\times 10^{-3}, \delta=8.94^{+1.96}_{-1.97}\times 10^{-3}$ are consistent with the standard slow-roll inflation. The dimensionless NC parameter $c_1$, appearing in the realisation of $\kappa$-Minkowski coordinates, also exhibits a Gaussian like distribution and is constrained to $c_1=4.79^{+1.01}_{-1.03}\times 10^{-3}$. More importantly, this analysis yields a constraint on the $\kappa$-deformation length scale as $\lambda=1.93^{+1.88}_{-1.33}\times 10^{-29}m$ at $1\sigma$ confidence level.

Our MCMC analysis using ACT DR6 data provides constraints on the \(\kappa\)-deformed space-time. We find that the data are consistent with standard slow-roll inflation and allow a deformation parameter \(\lambda\) roughly six orders of magnitude larger than the Planck scale, along with a non-vanishing NC parameter $c_1$, appearing in the realisation approach.

Recently, the NC scale of Moyal space-time has been constrained to $\theta<1.36\times 10^{-32}m$ at $2\sigma$ CL via the Bayesian analysis of inflationary models using Planck PR4 data \cite{Gandhi:2026ktq}. In contrast, our MCMC analysis of the $\kappa$-deformed inflationary model, employing ACT DR6 data, places an upper limit of $\lambda<6.03\times 10^{-29}m$ at $2\sigma$ CL. Although the order-of-magnitude difference between these bounds arises from distinct physical origins of these two NC frameworks, both constraints suggest that current CMB observations could probe Planckian and sub-Planckian scales. Future datasets from next-generation CMB experiments could further refine these bounds, potentially revealing the first signatures of quantum space-time structures.

Using the best-fit values from our MCMC analysis, we obtain $|\Delta_{\kappa}| \lesssim 2.07\%$ (where $\Delta_{\kappa} \equiv \Delta\mathcal{P}_{\mathcal{R}}/\mathcal{P}_{\mathcal{R}} \simeq \frac{16}{9}\lambda^2 c_1^2 H^2 \ln^2 k$) and $|\Delta n_s| \lesssim 0.018$ (where $\Delta n_s \simeq \frac{32}{9}\lambda^2 c_1^2 H^2 \ln k\,(1-\epsilon\ln k)$), at the CMB damping scales $k\sim 0.1 - 1\,\text{Mpc}^{-1}$. Further, these corrections induce shift in $S_8$ and $H_0$ measurements by $\Delta S_8 \sim 0.00824$ and $\Delta H_0 \sim 0.909\,\text{km}\,\text{s}^{-1}\text{Mpc}^{-1}$. However, these enhancements will only reduce the $S_8$ and $H_0$ tensions to $2.23\sigma$ and $4.20\sigma$, in comparison to their current values $2.72\sigma$ and $5\sigma$, respectively. Hence the $\kappa$-deformation cannot solve the current $S_8$ and $H_0$ tensions \cite{Sabogal:2024yha,Pantos:2026koc} in cosmology, although the $\kappa$ corrections slightly suppresses them.
     
\section{Discussion and Conclusion}\label{sec5}
    
In this work, we have investigated the imprints of $\kappa$-Minkowski NC space-time on the primordial perturbations generated during inflation. Employing the $\kappa$-deformed star product formalism, we have constructed the bilinear action for the curvature perturbation and derived the corresponding modified Mukhanov–Sasaki equation in $\kappa$-deformed space-time. In order to avoid the Ostrogradsky instability associated with the higher-order time derivative terms introduced by the star product, we have adopted an effective field theory approach, consistently eliminating these terms at the level of the equations of motion up to first order in the deformation parameter $\lambda$.

We obtain the perturbative solutions to the modified Mukhanov-Sasaki equation, in  the super-horizon and sub-horizon limits separately, and calculated the $\kappa$-deformed power spectrum to the curvature perturbation at the horizon crossing. Here the NC corrections to the primordial power spectrum is induced by the star product introduced in the bilinear action, in contrast to \cite{Rajagopal:2025vbs}, where the $\kappa$-deformations to the power spectrum have appeared from the $2$-point correlator via $\kappa$-deformed oscillator algebra. The leading-order correction to the scalar power spectrum contains a $(\ln k)^2$ dependence, showing that $\kappa$-deformed non-commutativity induces scale dependence. Moreover, this $\kappa$-deformed power spectrum preserves both parity and rotational invariance, which is a direct consequence of the fact that the $\kappa$-Minkowski space-time structure respects rotational symmetry, although it deforms Lorentz symmetry, and hence the resulting power spectrum becomes direction independent, thereby maintaining the statistical isotropy. 
    
The \(\kappa\)-deformed power spectrum for tensor perturbations (see Eq.(\ref{power-tensor})) also acquires the same correction factor as the curvature perturbation, since the $\kappa$-star product deforms the Mukhanov–Sasaki equation for both curvature and tensor perturbations in an equivalent manner. Consequently, the leading order correction factors for the scalar and tensor perturbations remain same, and as a result the scalar-to-tensor ratio (i.e., $r=\frac{\mathcal{P}_{\mathcal{R}}}{\mathcal{P}_T}=16\varepsilon$) is unaffected under the $\kappa$-deformation. This result is in agreement with that obtained in \cite{Rajagopal:2025vbs}, using the $\kappa$-deformed oscillator algebra.

The scalar spectral index also has an explicit dependence on $\ln k$, which persists even when the slow-roll parameters remain constant. The scale dependence of the spectral index provides a distinctive observational signature of $\kappa$-deformed space-time. Hence, we performed a Bayesian MCMC analysis using ACT DR6 data to constrain the model parameters, which yields the results \(\epsilon = 1.02^{+0.498}_{-0.460}\times 10^{-3}\) and \(\delta =8.94^{+1.96}_{-1.97}\times 10^{-3}\), consistent with standard slow-roll inflation.  More importantly, we constrain the dimensionless NC parameter to \(c_1 = 4.79^{+1.01}_{-1.03}\times 10^{-3}\) at $1\sigma$ CL. and \(\kappa\)-deformation length scale to $\lambda=1.93^{+1.88}_{-1.33}\times 10^{-29}m$ at \(1\sigma\) CL. This bound is approximately four orders of magnitude larger than the Planck scale, indicating that current CMB observations are sensitive to quantum spacetime structures at scales significantly above the Planck length. However from this best fit values, we find that the $\kappa$-deformed scale dependent enhancement is insufficient to address the current $S_8$ and $H_0$ tensions in cosmology \cite{Sabogal:2024yha,Pantos:2026koc}. Thus the upcoming CMB experiments with improved sensitivity to the running of the spectral index could provide some empirical evidence for quantum spacetime structures.

Extending this analysis by deriving the full non-perturbative \(\kappa\)-deformed gravitational action without matching approximations, and exploring the associated non-Gaussianities in CMB polarization, would provide a more complete picture of quantum gravitational effects in the early universe.

\section*{Acknowledgment}

This research work is supported by the National Natural Science Foundation of China (Grant No.~12275080) and the Innovative Research Group of Hunan Province (Grant No.~2024JJ1006).

\begin{appendix}
\section{}

The standard expression of the bilinear action for the tensor perturbation is \cite{Baumann:2022mni}
\begin{equation}\label{bilinear-tensor}
    S^{(2)}=\frac{1}{2} \int d^{4} x ~\sqrt{-g} ~g^{\mu \nu}\partial_{\mu} h_{ij}\cdot\partial_{\nu}h^{ij},
\end{equation}
where $h_{ij}$ is the transverse traceless tensor perturbation. We incorporate the NC deformations into this action using the star product formalism as discussed in Section.(\ref{sec2})
\begin{equation}\label{bilinear-tensor-star}
    S^{(2)}=\frac{1}{2} \int d^{4} x ~\sqrt{-g} ~g^{\mu \nu}\partial_{\mu} h_{ij}\star\partial_{\nu}h^{ij}.    
\end{equation}
Following the procedure as shown in Section.(\ref{sec2}), we the $\kappa$-deformed Mukhanov-Sasaki equation for the tensor perturbation modes as
\begin{equation}\label{ms-tensor}
		h_{k}^{\prime \prime}+\left(k^{2}-\frac{a^{\prime \prime}}{a}\right) h_{k}+i \lambda\left(c_{1} \partial_{\eta}^{2}\left(\frac{\eta}{a} h_{k}^{\prime \prime}\right)+c_{2} k^{2} \partial_{\eta}\left(\frac{\eta}{a} h_{k}^{\prime}\right)+\frac{\eta}{a} k^{4} h_{k}\right)=0,
	\end{equation}
where $h_k$ is the Fourier transform of the tensor perturbation modes $h_{ij}$, which has two independent polarisation states $(+,\times)$. Solving this equation, as discussed in Section.(\ref{sec3}), we get the explicit solution at horizon crossing and obtain $|h_k|^2=\frac{2 H^{2}}{k^3}\left(1+\frac{16}{9} \lambda^{2} c_{1}^{2} H^{2} \ln ^{2} k\right)$. Considering two polarisation modes, we get the tensor power spectrum using the standard expression $\mathcal{P}_T=\frac{k^3}{2\pi^2}2\cdot|h_k|^2$, as
\begin{equation}\label{power-tensor}
 \mathcal{P}_T= \frac{2 H^2}{\pi^2}\left(1+\frac{16}{9} \lambda^{2} c_{1}^{2} H^{2} \ln ^{2} k\right)
\end{equation}
This correction factor is exactly the same on obtained for the curvature perturbation, since the star product induces equivalent deformations to the scalar and tensor perturbations in the bilinear action.

\end{appendix}

	\bibliography{reference}

\end{document}